\documentclass{article}
\usepackage{epsfig}

\newtheorem{thm}{Theorem}
\newtheorem{lem}{Lemma}

\newtheorem{fig}{Figure}

\def\ligne#1{\hbox to \hsize{#1}}
\def\PlacerEn#1 #2 #3 {\rlap{\kern#1\raise#2\hbox{#3}}}

\font\itix=cmti9

\def\leurre{\noindent\leftskip0pt\small\baselineskip 10pt}

\begin{document}
\title{The Domino Problem of the Hyperbolic Plane Is Undecidable}
\author{Maurice Margenstern,\\
        Universit\'e Paul Verlaine $-$ Metz,\\
        LITA, EA 3097, IUT de Metz,\\
        \^Ile du Saulcy,\\
        57045 METZ C\'edex, FRANCE,\\
        {\it e-mail: margens@univ-mezt.fr}
       }
\maketitle

\begin{abstract}
In this paper, we prove that the general tiling problem of the hyperbolic plane
is undecidable by proving a slightly stronger version using only a regular 
polygon as the basic shape of the tiles. The problem was raised by a paper of 
Raphael Robinson in 1971, in his famous simplified proof that the general tiling
problem is undecidable for the Euclidean plane, initially proved by Robert Berger
in~1966.
\end{abstract}

\section{Introduction}

   The question, whether it is possible to tile the plane with copies of a fixed
set of tiles was raised by Wang, \cite{wang} in the late 50's of the
previous century. Wang solved the {\it origin-constrained} problem which consists 
in fixing an initial tile in the above finite set of tiles. Indeed, fixing one 
tile is enough to entail the undecidability of the problem. The general case, later
called the {\bf general tiling problem} in this paper, $GTP$ in short, without 
condition, in particular with no fixed initial tile, was proved undecidable 
by Berger in 1966, \cite{berger}. Both Wang's and Berger's proofs deal with the 
problem in the Euclidean plane. In 1971, Robinson found an alternative, simpler 
proof of the undecidability of the general problem in the Euclidean plane, 
see \cite{robinson1}. In this 1971 paper, he raises the question of the general 
problem for the hyperbolic plane. Seven years later, in 1978, he proved that 
in the hyperbolic plane, the origin-constrained problem is undecidable, 
see \cite{robinson2}. Up to now, $GTP$ remained open.

   In this paper, we give a synthetic presentation of the techniques contained 
on several technical papers deposited on {\it arXiv}, see \cite{mmarXiv2,%
%periodic domino:
mmarXiv5}, and on the web site of the author, in particular \cite{mmnewtechund}.

   In the second section, we remind a few features of all proofs of this problem.
Then, we turn to the hyperbolic case, assuming that the reader is familiar with
hyperbolic geometry, at least with its popular models: Poincar\'e's disc or 
half-plane.
We refer the reader to~\cite{mmbook1} and to~\cite{mmDMTCS} for preliminaries and
other bibliographical references.

   In the third section, we sketchilly present the frame of the construction. 
In the fourth section, we present a needed interlude, a parnthesis on brackets,
which is a basic ingredient of the proof. In the fifth section, we lift up 
this line construction to a planar Euclidean one. In the sixth section, we show 
how to implement the Euclidean construction in the hyperbolic plane, using the
specific properties indicated in the third section. In the seventh section,
we complete the proof of the main result:

\begin{thm}\label{undec}
\it The domino problem of the hyperbolic plane is undecidable.
\end{thm}

   From theorem~1, we immediately conclude that $GTP$ is undecidable in the
hyperbolic plane. 

   In the eighth section, we give several corollaries of the construction
and we conclude in two directions. The first one wonders whether it is possible 
to simplify the construction. The second direction tries to see what might be 
learned from the construction leading to theorem~\ref{undec}.

   Before turning to section~2, let us remark that an alternative proof of~$GTP$
is claimed by Jarkko Kari, see~\cite{jkariDavidson,kariMCU}. His proof is 
completely different of this one. It is completely combinatoric and it makes 
use of a non-effective argument. Here, we have no room to discuss this latter 
point.

\section{The general strategy}

In the proofs of $GTP$ in the Euclidean plane by Berger and Robinson, there is 
an assumption which is implicit and was, most probably, considered as obvious at
that time.   

   Consider a finite set~$S$ of {\bf prototiles}. We call {\bf solution} of the
tiling of the plane by~$S$ a partition~$\cal P$ such that the closure of any
element of~$\cal P$ is a copy of an element of~$S$. We notice that this 
definition contains the traditonal condition on matching signs in the case when 
the elements of~$S$ possess signs.

   Note that $GTP$ can be formalized as follows:
\vskip 3pt
\ligne{\hfill$\forall S\ \ (\exists\,{\cal P}\ sol({\cal P},S))\vee%
              \neg(\exists\,{\cal P}\ sol({\cal P},S)),$
\hfill}
\vskip 3pt
\noindent
where $\vee$ is interpreted in a constructive way: there is an algorithm
which, applied to~$S$ provides us with~'yes' if there is a solution and~'no'
if there is none.

   The origin-constrained problem can be formalized in a similar way by:
\vskip 3pt
\ligne{\hfill$\forall (S,a)\ \ (\exists\,{\cal P}\ sol({\cal P},S,a))\vee%
              \neg(\exists\,{\cal P}\ sol({\cal P},S,a)),$
\hfill}
\vskip 3pt
\noindent
where $a\in S$, with the same algorithmic interpretation of~$\vee$.

   Now, note that if we have a solution of~$GTP$, wa also have a solution of 
the origin-constrained problem, with the facility that we may choose the
first tile. However, to prove that~$GTP$ has no solution, we have to prove
that, whatever the initial tile, the corresponding origin-constrained problem
also has no solution.

   However, Berger's and Robinson's proofs consider that we start the
construction with a special tile, called the {\bf origin}. There is no
contradiction with what we have just said, because they force the tiling to
have a dense subset of origins. In the construction, the origins start the
simulation of the space-time diagram of the computation of a Turing machine~$M$.
All origins compute the same machine~$M$ which can be assumed to start from
an empty tape. The origins define infinitely many domains of computation of
infinitely many sizes. If the machine does not halt, starting from an origin,
it is possible to tile the plane. If the machine halts, whatever the initial 
tile, we nearby find an origin and, from this one, we shall eventually fall
into a domain which contains the halting of the machine: at this point, it
is easy to prevent the tiling to go on.

   The present construction aims at the same goal.

\section{The general frame: the tiling $\{7,3\}$ and its mantilla}

Our construction takes place in a particular tiling of the hyperbolic plane:
the tessellation $\{7,3\}$ which we call the {\bf ternary heptagrid}, simply 
{\bf heptagrid}, for short, see \cite{ibkmACRI,mmJCA,mmbook1}. It is generated
by the regular heptagon with vertex angle~$\displaystyle{{2\pi}\over3}$ by
reflection in its sides and, recursively, by reflection of the images in their
sides. The background of figure~\ref{til_mantilla} goves an illustration of this
tiling in the Poincar\'es disc.

\vskip 5pt
\setbox110=\hbox{\epsfig{file=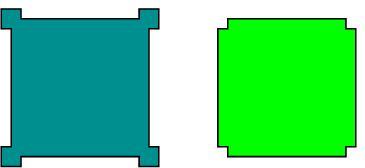,width=150pt}}
\setbox112=\hbox{\epsfig{file=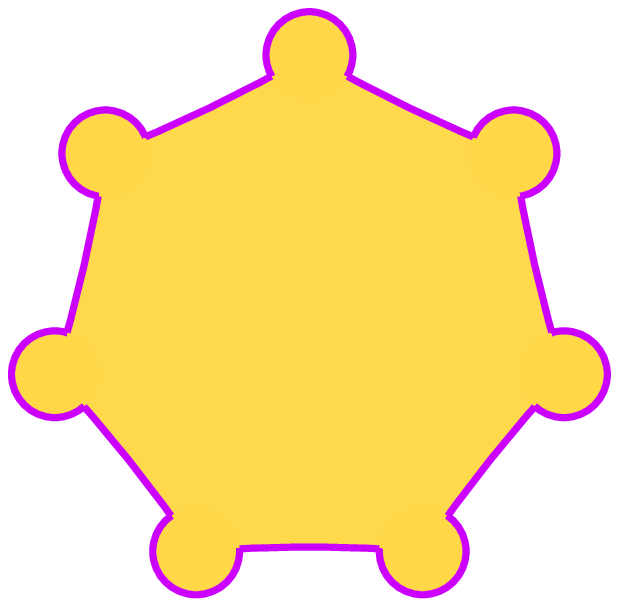,width=130pt}}
\setbox114=\hbox{\epsfig{file=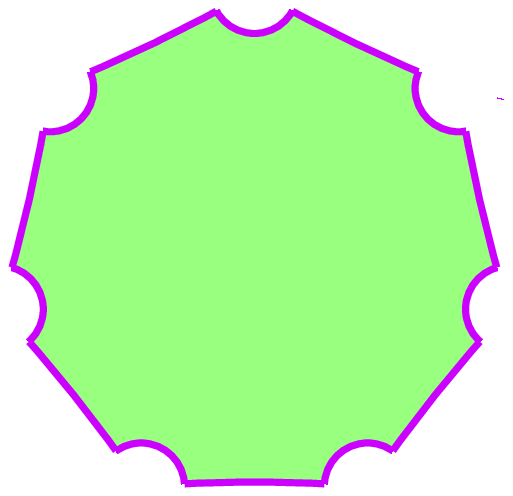,width=130pt}}
\vtop{
\ligne{\hfill
\PlacerEn {-160pt} {0pt} \box110
\PlacerEn {-20pt} {0pt} \box112
\PlacerEn {60pt} {0pt} \box114
\PlacerEn {80pt} {10pt} {$a$}
\PlacerEn {160pt} {10pt} {$b$}
\hfill}
\vskip-15pt
\begin{fig}\label{pseudo_robinson}
\leurre
On the left: Robinson's basic tiles for the undecidability of the tiling problem
in the Euclidean case. On the right: the tiles~$a$ and~$b$ are a 'literal'
translation of Robinson's basic tiles to the situation of the ternary heptagrid.
\end{fig}
}
   
   We consider a special tiling based on the tessellation $\{7,3\}$ which is 
motivated b the following consideration. In figure~\ref{pseudo_robinson}, the
left-hand side indicates the basic shapes of the tiles devised by Robinson in 
the construction of the tiling used in his proof of the undecidability 
of~$GTP$. The right-hand side of the figure gives the 'literal' translation
of these tiles for the heptagrid. It is not difficult to see that it
is not possible to tile the hyperbolic plane with the tiles~$a$ and~$b$. However,
a slight modification of the tile~$b$, see the tile~$c$ of 
figure~\ref{pseudo_momo}, leads to the solution. 

\vskip 5pt
\setbox110=\hbox{\epsfig{file=pseudo_robinson_a.ps,width=130pt}}
\setbox112=\hbox{\epsfig{file=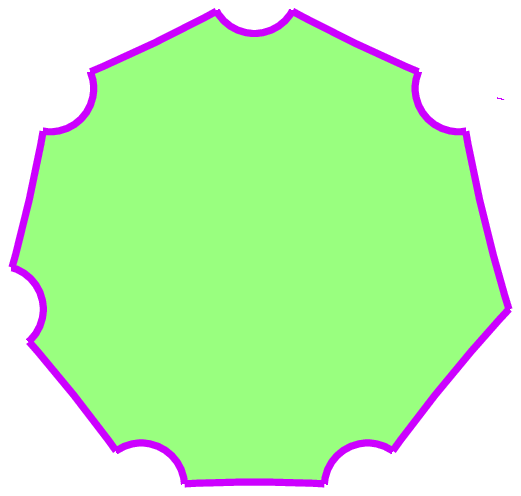,width=130pt}}
\setbox114=\hbox{\epsfig{file=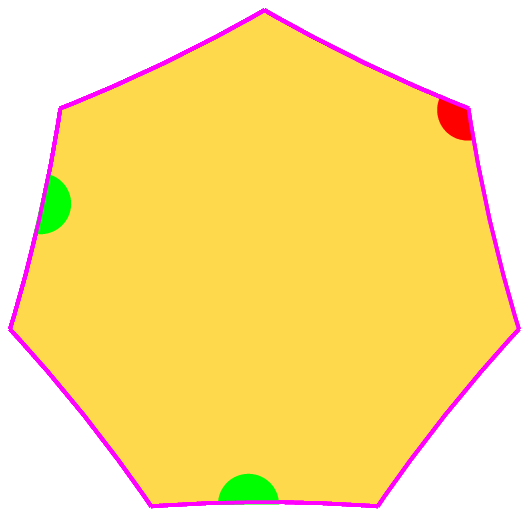,width=130pt}}
\setbox116=\hbox{\epsfig{file=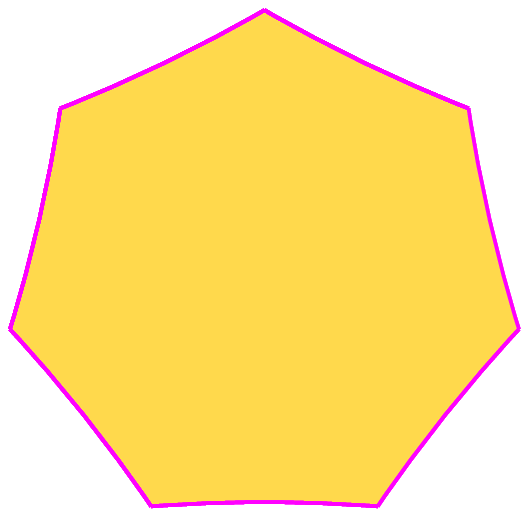,width=130pt}}
\vtop{
\ligne{\hfill
\PlacerEn {-190pt} {0pt} \box110
\PlacerEn {-105pt} {0pt} \box112
\PlacerEn {-20pt} {0pt} \box114
\PlacerEn {60pt} {0pt} \box116
\PlacerEn {-90pt} {10pt} {$a$}
\PlacerEn {-10pt} {10pt} {$c$}
\PlacerEn {80pt} {10pt} {$\alpha$}
\PlacerEn {160pt} {10pt} {$\beta$}
\hfill}
\vskip-15pt
\begin{fig}\label{pseudo_momo}
\leurre
On the left: change in the tiles \`a la Robinson. On the right: their
translation in pure Wang tiles.
\end{fig}
}   

   On the right-hand side of figure~\ref{pseudo_momo}, we have the translation
of the tiles~$a$ and~$c$ into genuine {\it \`a la} Wang tiles. The 
pattern~$\alpha$ corresonds to~$a$ and the pattern~$\beta$ corresponds to~$c$.

\subsection{The mantilla}

   In the ternary heptagrid, a {\bf ball} of {\bf radius}~$n$ around a tile~$T_0$
is the set of tiles which are within distance~$n$ from~$T_0$ which we call the
{\bf centre} of the ball. The {\bf distance} of a tile~$T_0$ to another~$T_1$
is the number of tiles constituting the shortest path of adjacent tiles 
between~$T_0$ and~$T_1$. We call {\bf flower} a ball of radius~1.

   Now, the tiles~$\alpha$ and~$\beta$ of figure~\ref{pseudo_momo} can be 
assembled in flowers only: a tile~$\alpha$, which we call {\bf centre}, 
require to be surrounded by tiles~$\beta$ only, which we call {\bf petals}.
This is obtained by numbering the edges of the tiles~$\alpha$ from~1 to~7,
see \cite{mmtechund,mmnewtechund}. Now, a petal belongs to three flowers at
the same time by the very definition of the implementation. From this, there is 
a partial merging of the flowers. By definition, the resulting tiling is
the {\bf mantilla}.

\vskip 15pt
\setbox110=\hbox{\epsfig{file=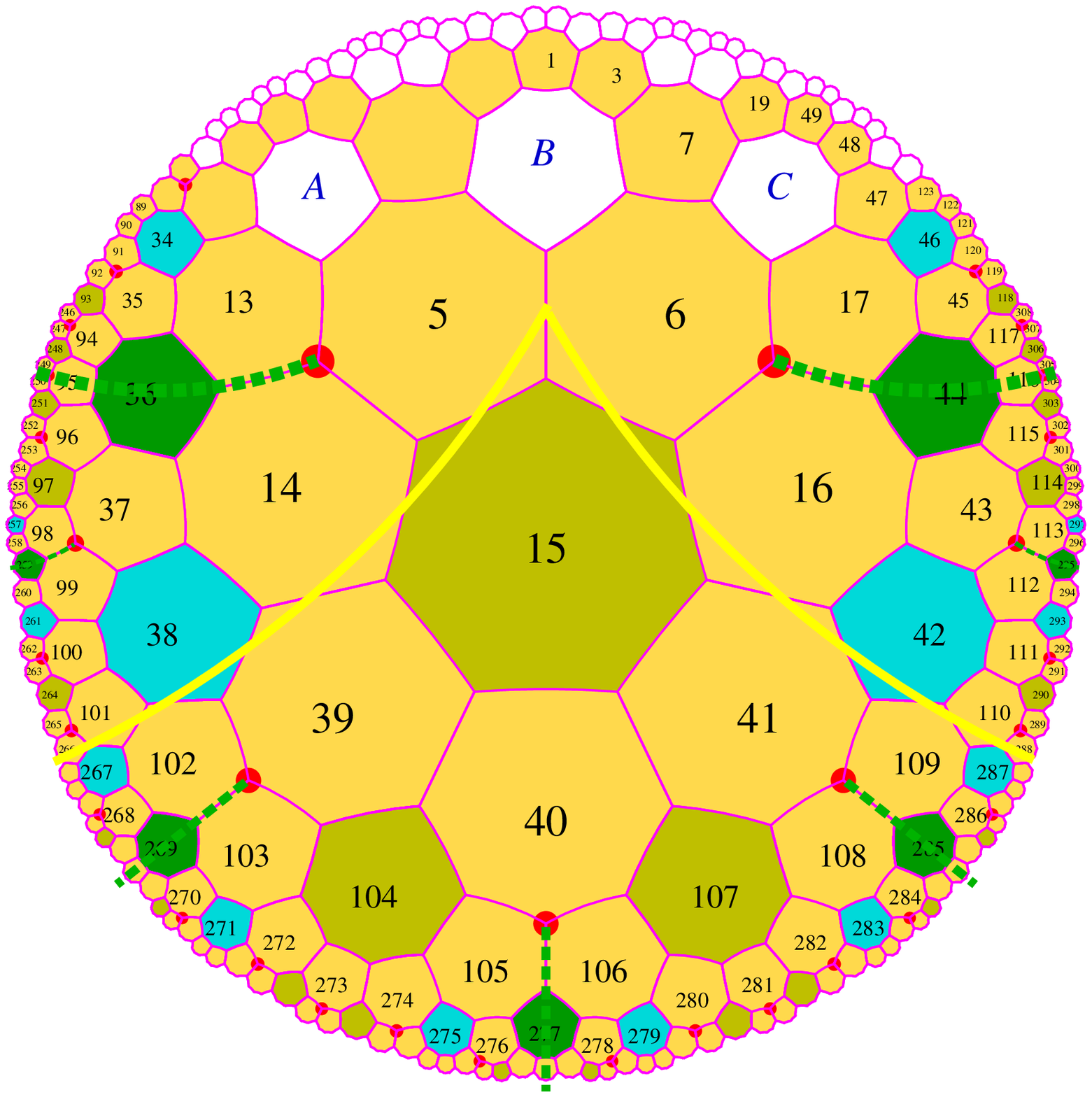,width=100pt}}
\setbox112=\hbox{\epsfig{file=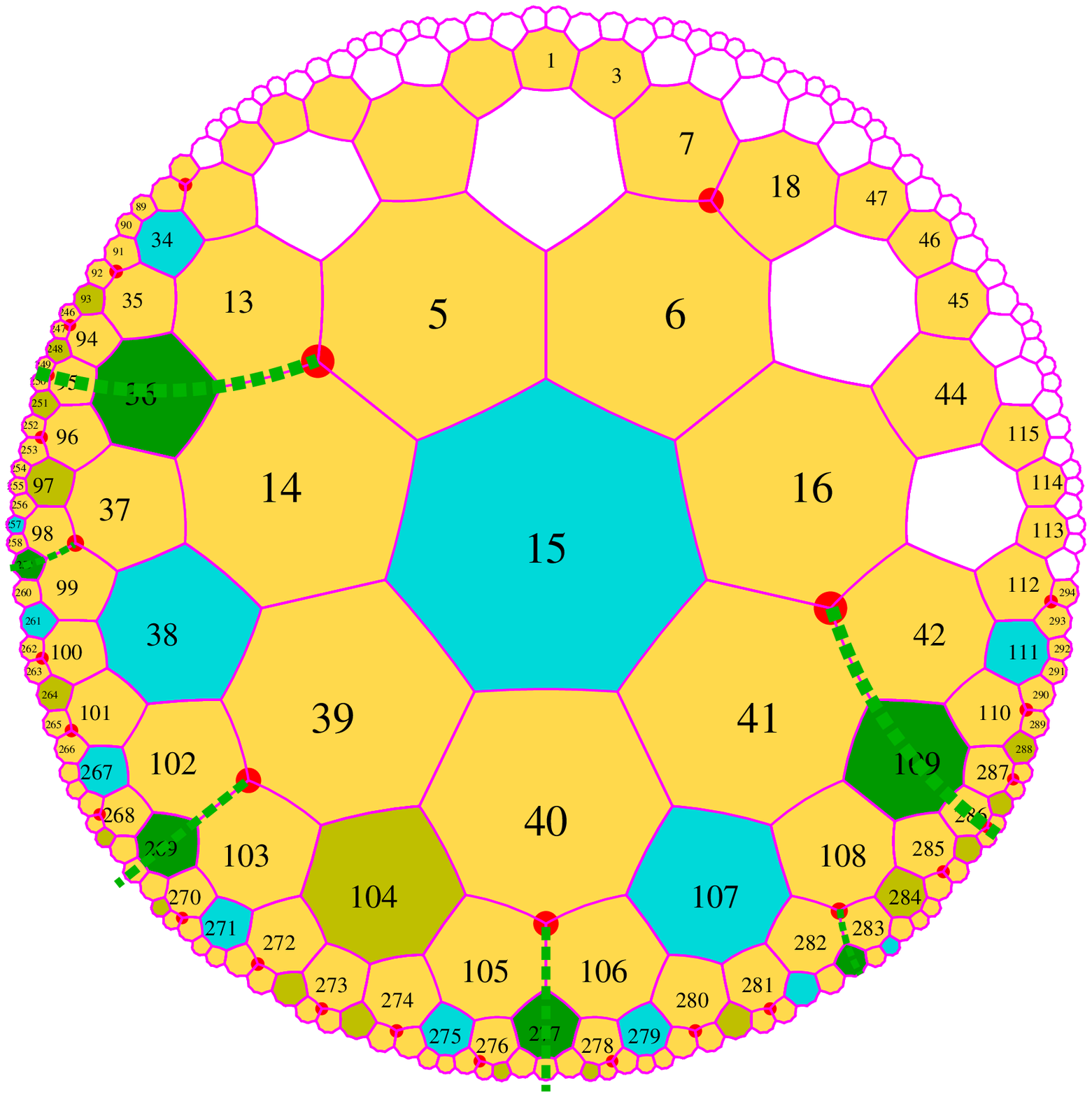,width=100pt}}
\setbox114=\hbox{\epsfig{file=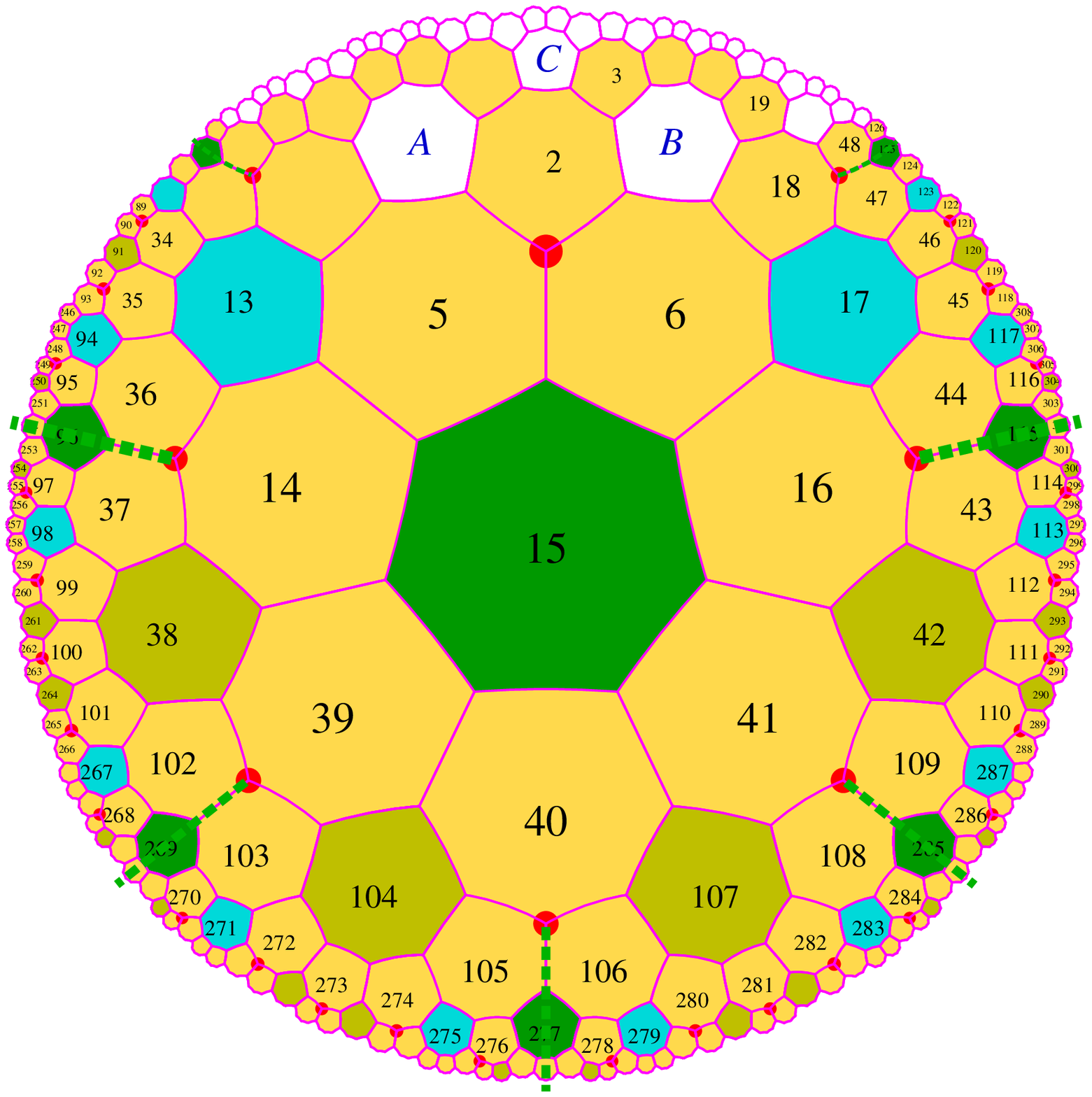,width=100pt}}
\vtop{
\ligne{\hfill
\PlacerEn {-345pt} {-60pt} \box110
\PlacerEn {-255pt} {-55pt} {$F$}
\PlacerEn {-225pt} {-60pt} \box112
\PlacerEn {-135pt} {-55pt} {$G$}
\PlacerEn {-105pt} {-60pt} \box114
\PlacerEn {-15pt} {-55pt} {\bf 8}
}
\begin{fig}\label{til_mantilla}
\leurre
Splitting of the sectors defined by the flowers. From left to right:
an $F$-sector, a $G$-sector and an {\bf 8}-sector.
\end{fig}
}

   It is not difficult to see tha there can be several types of flowers, 
considering the number of red vertices for which the other end of an edge is 
a vertex of a centre. We refer the reader to~\cite{mmtechund} for the
corresponding properties. Here, we simply take into consideration that we 
have three basic patterns of flowers, which we call $F$-, $G$- and 
{\bf 8}-flowers respectively. They are represented by figure~\ref{til_mantilla}.

   The figure also represents the way which allows to algorithmically construct
the mantilla. It consists in splitting the {\bf sectors} generated by each kind
of flowers in sub-sectors of the same kind and only them, which we call the
{\bf sons} of the flowers. From this, we easily devise a way to recursively
define a tiling. The construction is deterministic below the flower, and it
is non-deterministic when we proceed upwards. We do not make the notion of
top and bottom more precise: it will be done later. The exact description
of the splitting can be found in \cite{mmtechund}. We simply remark that 
such a splitting is an application of the general method described 
in~\cite{mmbook1,mmDMTCS}, for instance.

   Based on these considerations, we have the following result which is
thoroughly proved in \cite{mmtechund}.

\begin{lem}\label{mantilla_tiles} 
There is a set of $4$~tiles of type~$\alpha$ and $17$~tiles of type~$\beta$
which allows to tile the hyperbolic plane as a matnilla. Moreover, there is 
an algorithm to perform such a construction.
\end{lem}

\subsection{Trees of the mantilla}

Note that the leftmost flower of figure~\ref{til_mantilla}, which represents
an $F$-sector, also indicates a region delimited by continuous lines, yellow
in coloured figures. These lines are {\bf mid-point} lines, which pass through 
the mid-points of consecutive edges of heptagons of the heptagrid. As shown
in~\cite{ibkmACRI,mmJCA}, they delimit a Fibonacci tree. Let us remind that a
Fibonacci tree has two kinds of nodes: black ones and white ones. A black node
has two sons, a black and a white one. A white node has three sons, a black
and two white ones. In both cases, the black son is the leftmost son. The root
of a Fibonacci tree is a white node. The tiles inside the tree which are cut by 
these mid-point rays are called the {\bf borders} of the tree, while the set of
tiles spanned by the Fibonacci tree is called the {\bf area} of the tree.

   Say that an $F$-son of a $G$-flower is a {\bf seed} and the tree, rooted at
a seed is called a {\bf tree of the mantilla}. As the seeds are the candidates
for the construction of a computing region, they play an important r\^ole. From
figure~\ref{til_mantilla},, we can easily define the {\bf border} of a sector
which is a ray crossing \hbox{{\bf 8}-centres}. See \cite{mmtechund} for exact
definitions.

\begin{lem}\label{treesector}
The borders of a tree of the mantilla never meet the border of a sector.
\end{lem}

   From lemma~\ref{treesector}, as shown in \cite{mmtechund}, we easily obtain:

\begin{lem}\label{disjointtrees}
Consider two trees of the mantilla. Their borders never meet. Either their areas 
are disjoint or the area of one of them contains the area of the other.
\end{lem}

   From this, we can order the trees of the mantilla by inclusion of their areas.
It is clear that it is only a partial order. We are interested by the maximal
elements of this order. We call them {\bf threads}, see \cite{mmtechund} for an
exact definition. Threads are indexed by~$I\!\!N$ or~$Z\!\!\!Z$. We call
{\bf ultra-threads} the threads which are indexed by~$Z\!\!\!Z$. When there
are ultra-threads, two of them coincide, starting from a certain index. Note
that the union of the areas of the trees which belong to an ultra-thread is
the hyperbolic plane. There can be realizations of the mantilla with or without
ultra-threads.

\subsection{Isoclines}

In~\cite{mmnewtechund}, we have a new ingredient. We define the status of a tile
as~{\bf black} or~{\bf white}, defining them by the usual rules of such nodes 
in a Fibonacci tree. Then, we have the following property.

\begin{lem}\label{blackseed}
It is possible to require that {\bf 8}-centres are always black tiles. When
this is the case, a seed is always a black tile.
\end{lem}

\vskip 5pt
\setbox110=\hbox{\epsfig{file=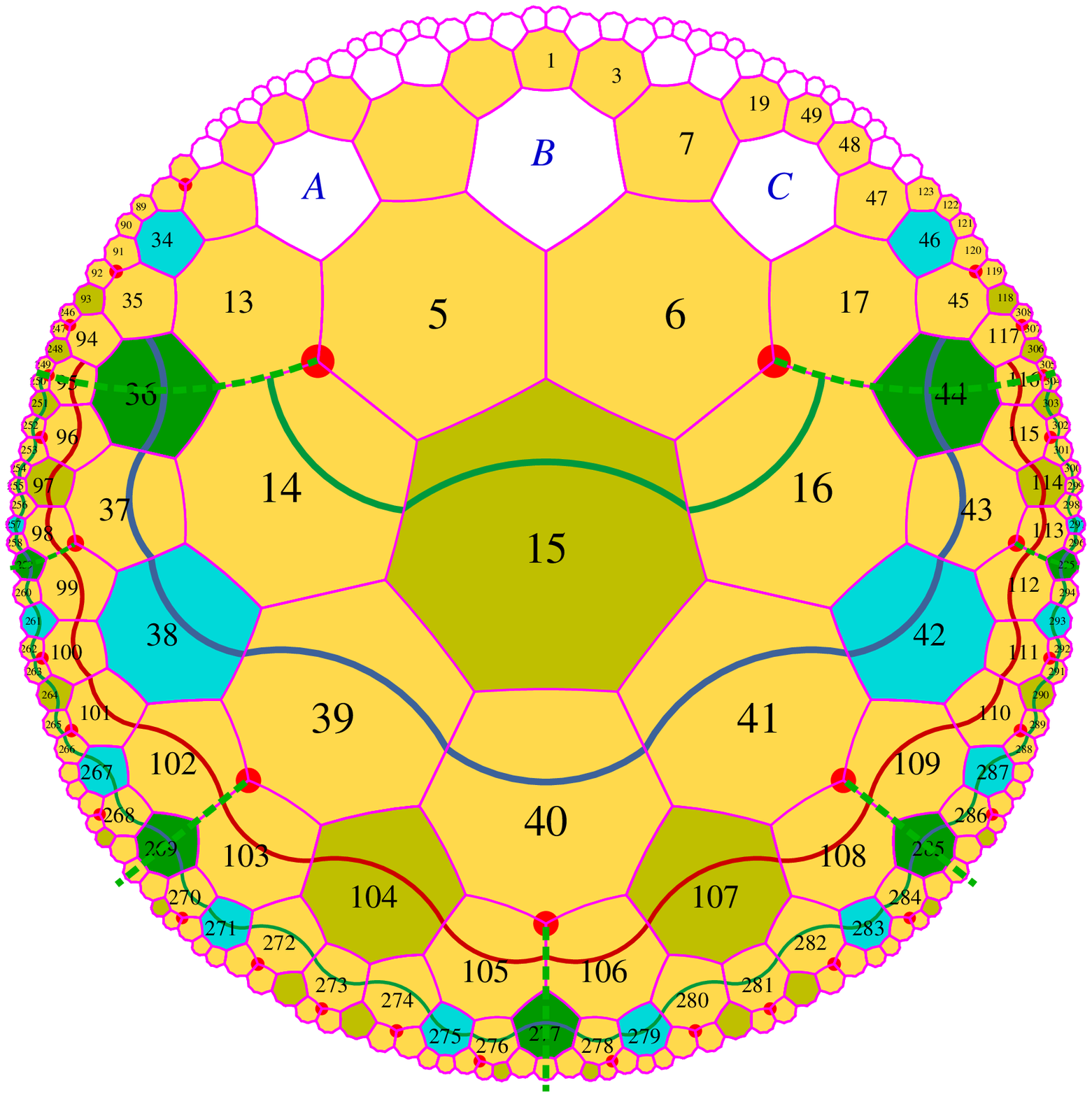,width=110pt}}
\setbox118=\hbox{\epsfig{file=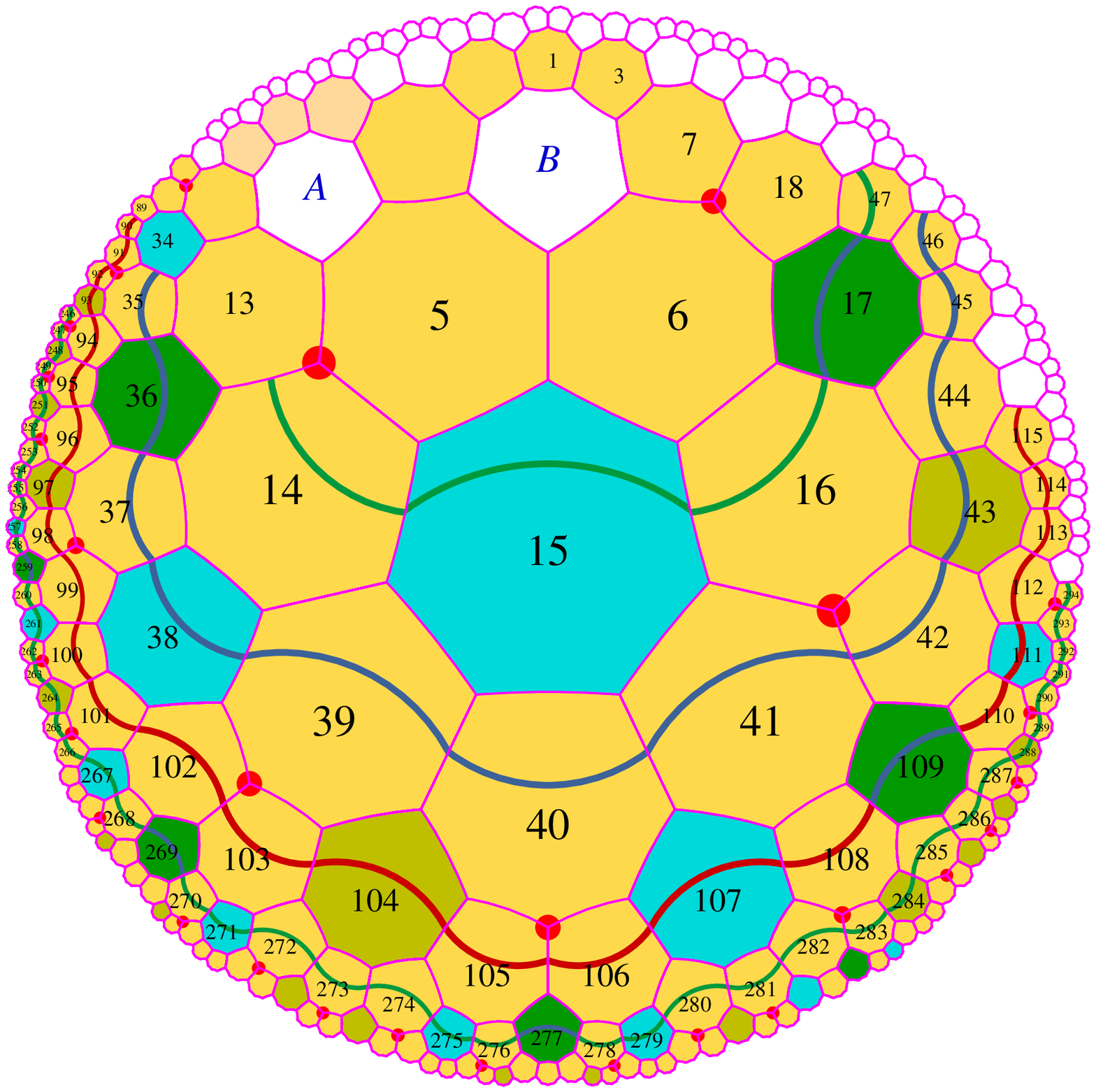,width=110pt}}
\vtop{
\ligne{\hfill
\PlacerEn {-310pt} {0pt} \box110
\PlacerEn {-150pt} {0pt} \box118
}
\begin{fig}\label{blackseedfig}
\leurre
The black tile property and the levels:
\vskip 0pt\noindent
On the left-hand side, a black $F$-centre; on the right-hand side,
a black $G_\ell$-centre. We can see the case of an {\bf 8}-centre on both
figures.
\end{fig}
}

   As shown in~\cite{mmnewtechund}, we can define arcs as follows: in a white
tile, the arc joins the mid-point of the sides which have a common vertex with 
the side shared by the father. In a black tile, the arc joins the mid-point of 
the sides shared by the father and the side shared by the uncle, which is on 
the left-hand side of the father. Joining arcs, we get paths. The maximal
paths are called {\bf isoclines}. They are illustrated on 
figure~\ref{blackseedfig}. An isocline is infinite and it splits the hyperbolic
plane into two infinite parts. The isoclines from the different trees match, 
even when the areas are disjoint.

\begin{lem}
\label{iso5}
Let the root of a tree of the mantilla~$T$ be on the isocline~$0$.
Then, there is a seed in the area of~$T$ on the isocline~$5$.
If an {\bf 8}-centre~$A$ is on the isocline~$0$, starting from the
isocline~$4$, there are seeds on all the levels. From the
isocline~$10$ there are seeds at a distance at most~$20$ from~$A$.
\end{lem}

   We number the isocline from 0 to 19 and repeat this, periodically.
This allows to give sense to {\bf upwards} and {\bf downwards} in
the hyperbolic plane.

\section{A parenthesis on brackets}

   We refer the reader to \cite{mmnewtechund} for an exact definition.
However, figure~\ref{silentfig}, below, illustrates the construction
which now, we sketchily describe.

   The generation~0 consists of points on a line which are regularly
spaced. The points are labelled $R$, $M$, $B$, $M$, in this order,
and the labelling is periodically repeated. An interval defined by
an $R$ and the next~$B$, on its right-hand side, is called
{\bf active} and an interval defined by a~$B$ and the next~$R$
on its right-hand side is called {\bf silent}. The generation~0 is
said to be {\bf blue}.

\vskip 5pt
\setbox110=\hbox{\epsfig{file=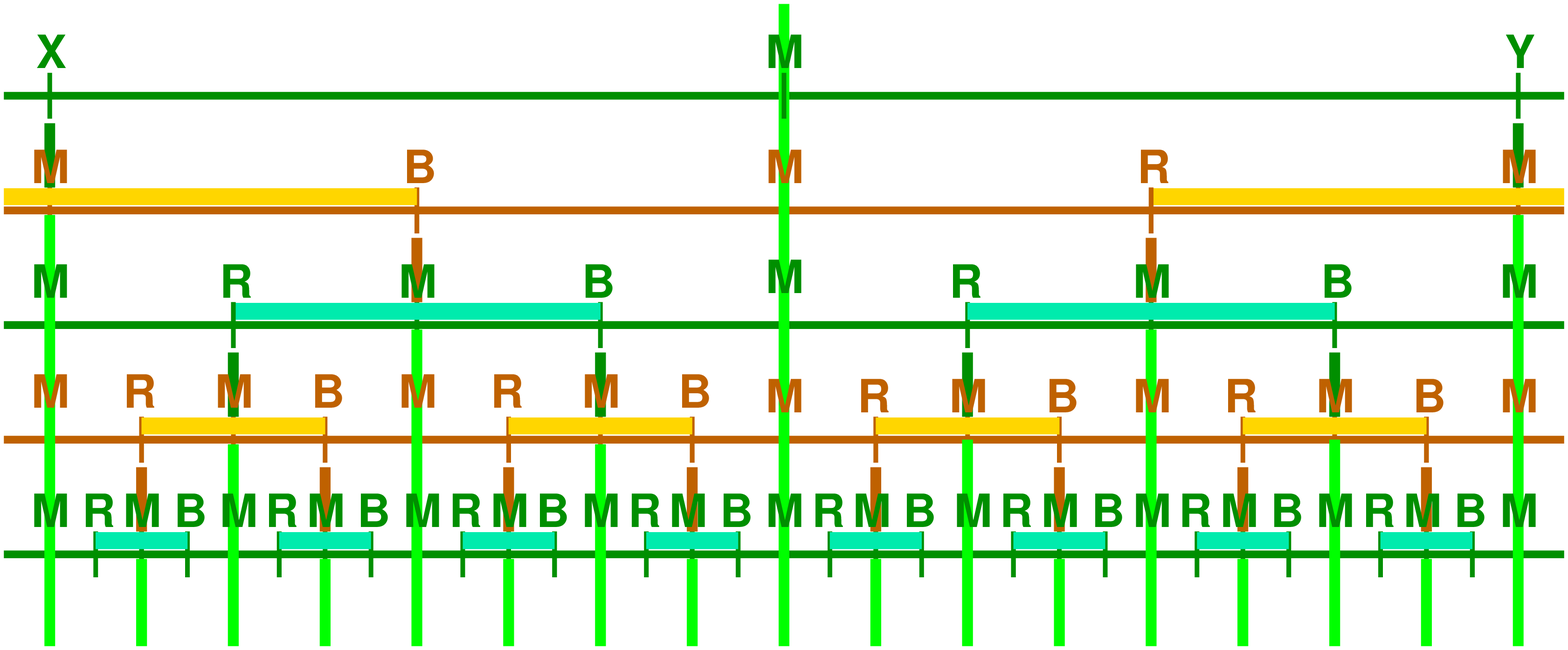,width=300pt}}
\vtop{
\ligne{\hfill
\PlacerEn {-327pt} {-10pt} \box110
}
\begin{fig}\label{silentfig}
\leurre
The silent and active intervals with respect to mid-point lines. The light green
vertical signals send the mid-point of the concerned interval to the next
generation. The colours are chosen to be easily replaced by red or blue inan 
opposite way. The ends~$X$ and~$Y$ indicate that the figure can be used
to study both active and silent intervals.
\end{fig}
}
\vskip 5pt
   Blue and red are said {\bf opposite}. Assume that the
generation~$n$ is defined. For the generation~$n$+1,
the points which we take into consideration are the points which are
still labelled~$M$ when the generation~$n$ is completed. Then, we
take at random an~$M$ which is the mid-point of an active interval
of the generation~$n$, and we label it, either~$R$ or~$B$. Next,
we define the active and silent intervals in the same way as for the
generation~0. The active and silent intervals of the generation~$n$+1
have a colour, opposite to that of the generation~$n$.

   When the process is achieved, we get an {\bf infinite model}.

   Deep results on the space of all these realizations are given by an
acurate analysis to be found in \cite{levin}. The interested reader should
have a look at this paper.

   For our purpose, infinte models have interesting properties, 
see \cite{mmnewtechund}. We cannot mention all of them
here. We postpone some of them to the Euclidean implementation with
triangles.

   Cut an infinite model at some letter and remove all
active intervals which contain this letter. What remains on the
right-hand side of the letter is called a {\bf semi-infinite model}.

   It can be proved that in a semi-infinite model, any letter~$y$
is contained in at most finitely many active intervals, see
\cite{mmnewtechund}.

\section{The interwoven triangles}

   Now, we lift up the active intervals as {\it triangles} in the Euclidean plane.
The triangles are isoceles and their heights are supported by the same line,
called the {\bf axis}, see figure~\ref{interwoven}.

   We also lift up silent intervals of the infinite model up to again
isoceles triangles with their heights on the axis. To distinguish them from 
the others, we call them {\bf phantoms}. We shall speak of {\bf trilaterals} for
properties shared by both triangles and phantoms.

\setbox110=\hbox{\epsfig{file=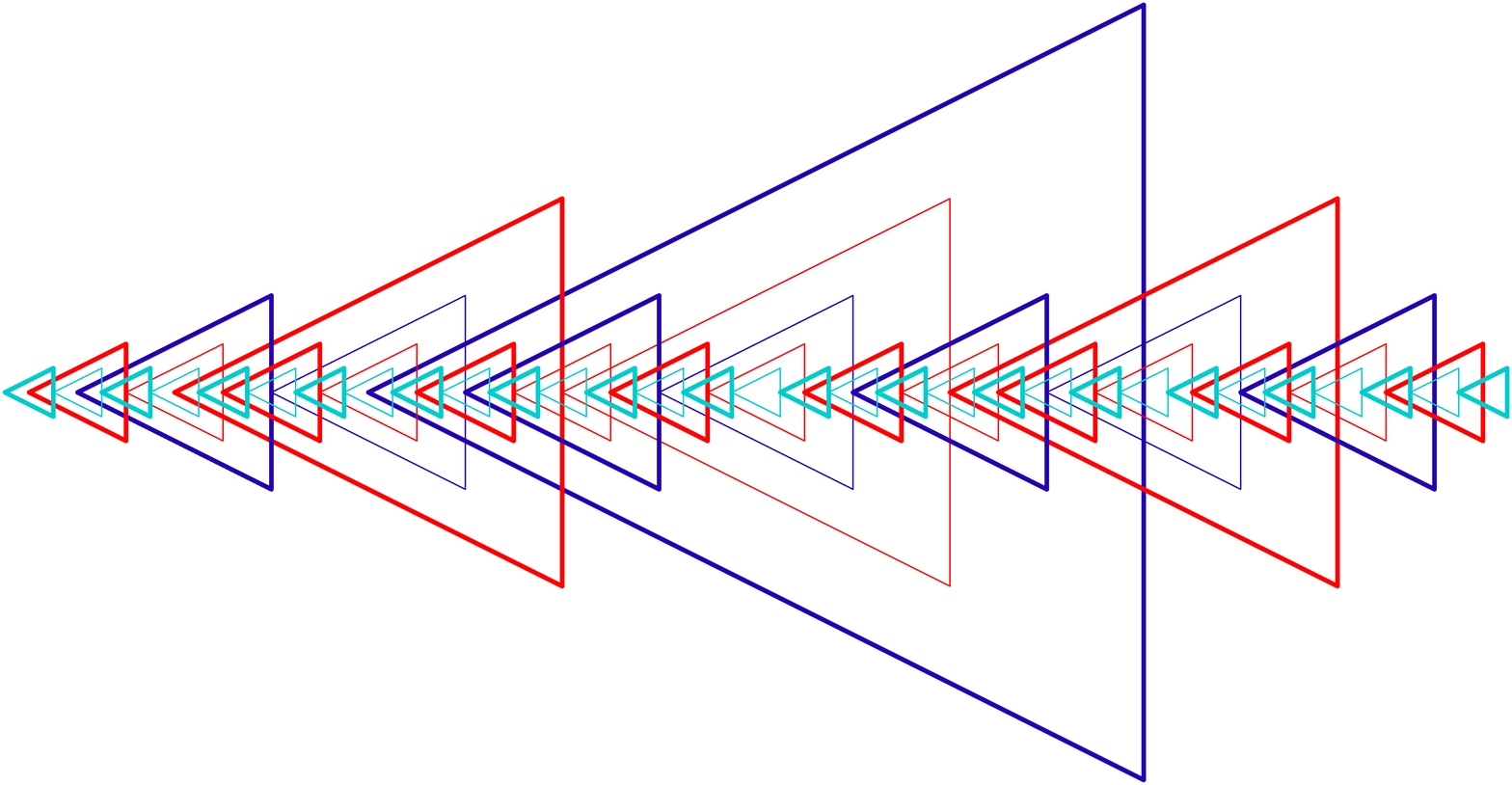,width=320pt}}
\vtop{
\ligne{\hfill
\PlacerEn {-350pt} {-90pt} \box110
}
\begin{fig}\label{interwoven}
An illustration of the interwoven triangles.
\end{fig}
}

   We have very interesting properties for our purpose.

\begin{lem}\label{towers}
Triangles of the same colour do not meet no overlap: they are disjoint or
embedded. Phantoms can be split into {\bf towers} of embedded phantoms with
the same mid-point and with alternating colours. Trilaterals can meet by a 
basis cutting the halves of the legs which contains the vertex.  
\end{lem}

   From the construction of the abstract brackets, we get a simple algorithm to 
construct the inteerwoven triangles.

   The generation~0 is fixed by alternating triangles and phantoms. The 
mid-distance lines of the phantoms of generation~0 grow a horizontal green
signal which crosses the legs of the phantom which they meet.

   When the generation~$n$ is completed, a vertex of a triangle or a phantom
is put on the intersection of the axis with the mid-distance line of a 
{\bf triangle} of generation~$n$. The legs grow until they meet a green signal,
traveling on a horizontal line. If the trilateral is a tirangle, the legs stop
the green signal. If the trilateral is a phantom, the green signal crosses the 
legs. In both cases, the legs go on, until thy meet a basis of their colour.
They stop it and constitute a trilateral of the generation~$n$+1. Now, the
intersection of the basis of the trilateral with the axis requires a vertex:
of a triangle, if the basis belongs to a phantom, of a phantom, if the basis
belongs to a triangle. The process is repeated endlessly.

   The detection of the green signal and of the correct basis are facilitated
by the following mechanism: the legs of red triangles emit horizontal signals
which have a laterality. A right-hand side signal emits a right-hand side
signal and a left-hand side leg emits a left-hand side signal. We do not 
provide a tile for the meeting of such signals inside a triangle. This allows
to detect what corresponds to the free letters and which we call the {\bf free 
rows} of a red triangle.

\begin{lem}\label{euclid_tiles}
The interwoven triangles can be obtained by a tiling of the Euclidean
plane which can be forced by a set of $190$~tiles.
\end{lem}

   In \cite{mmnewtechund}, we display the corresponding tiles which are 
in a square format, and we also describe them with the help of formulas 
taking into account the properties of lemma~\ref{towers}.

\section{Implementing the interwoven triangles in the hyperbolic plane}

   The idea of the implementation in the hyperbolic plane is based on the
following observation.

   From lemma~\ref{iso5}, we define the isoclines~0, 5, 10 and~15 to play the
r\^ole of the rows in the Euclidean implementation. The trilaterals will be
constructed on trees of the mantilla. A vertex will be a seed, and the legs 
are supported by the borders of the tree. The basis is defined by an isocline
which cuts both borders of the tree.

   As there are 6~seeds on the isocline~5 inside the tree defined by a seed
at the isocline~0, there are 6~trilaterals of the generation~1 raised by
a triangle of the generation~0. And so, contrarily to what happens in the 
Euclidean construction, we have several trilaterals of the same generation
for the same set of isoclines crossed by the legs of these trilaterals.

   Call {\bf latitude} of a trilateral the set of isoclines which are crossed
by its legs, vertex and basis being included.

   It is not difficult to see that there will be infinitely many trilaterals
within a given latitude. This requires to {\bf synchronize} the choice of
triangle or phantom when turing from one generation to the next one. Also,
we can imagine that horizontal signals coming from different triangles of 
the same latitude and with different lateralities will meet. This will 
require to tune many details in order to maintain the guidelines of the
algorithm, described in section~5.

   The idea will be to synchronize the bases, the vertices and the signals
on the mid-distance lines. We shall also have to see how the axis is
implemented.

   From what we have already seen, there is no moe one axis, but a lot of
them. In fact, what we called an axis can be materialized by a thread. AS
most threads are indexed by~$I\!\!N$ only, we have always the implementation 
of a semi-infinite model. Now, we shall manage the implementation in such 
a way that the semi-infinite models are simply different cuts of the same
infinite model. The possibility of the realization of the infinite model
in the case of an ultra-thread brings in no harm: it can be viewed as a
cut at infinity.

\subsection{The scent}   

   By definition, we decide that all seeds which are on an isocline~0
are {\bf active}, which means that they actually grow legs of a 
triangle of the generation~0. This is enough to guarantee that the
set of active seeds is dense in the hyperbolic plane. Next, an active
seed diffuses a {\bf scent} inside its trilateral until the fifth
isocline, starting from this seed, is reached. Seeds which receive the
scent, and only them, become active. An active seed triggers the green 
signal when it reaches an isocline~5 or an isocline~15. By construction, 
the generation~0 is not determined by the meeting of a green signal. But 
the others are. 

   We can see that the scent process constructs a tree. The branches of the
tree materilize the thread which implements the considered semi-infinite 
models. Note that the above synchronizatoin mechanism fixes things for
spaces between triangles but also inside them.

\subsection{Horzontals signals with a laterality}

   Due to the occurrene of several trilaterals within a given latitude,
now we have to require that all triangles, both red and blue and blue~0,
emit a signal along their legs. The signal will be called {\bf upper}
when it is emitted by the legs, but there is an exception: the
vertices do not emit any horizontal signal. There is another exception:
the corners of a phantom, as well as those of a triangle, also emit an
upper signal. The colour and the laterality of this signal are those of
the corner.

   Now, in between two contiguous triangles of the same latitude, horizontal
signals of the same colour but with a different laterality will meet. We
have to allow such a meeting, which will be performed by an appropriate tile
which we call a {\bf join} tile. There is a join-tile for red and blue signals,
as well as for the orange signals. The join-tile, see the pattern represented
by~figure~\ref{the_join}, illustrates such a junction. On the left-hand side,
we have the right-hand side signal and, on the right-hand side, we have the
left-hand side one. We also require that an upper horizontal signal of a given 
laterality may cross a leg of a trialteral of the same colour, only if it has
the same laterality as the leg. Note that the opposite junction cannot be 
obtained,  as turing tiles is impossible in our setting. This impossibility is
guaranteed by the existence of the isoclines and thir numbering.

\begin{lem}\label{unilateral}
An upper horizontal signal with a laterality cannot join two legs of the same
tilateral.
\end{lem}

\noindent
Proof: easy corollary of the rule about the meeting of legs with an
upper horizonal signal. 

   Accordingly, the legs of a trilateral may only be joined by a horizontal
signal which has no laterality.

\setbox110=\hbox{\epsfig{file=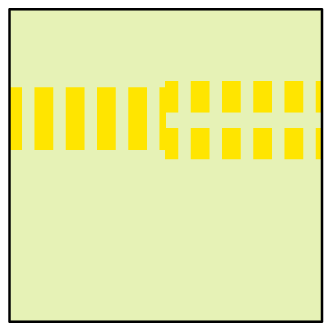,width=180pt}}
\vskip -30pt
\vtop{
\ligne{\hfill
\PlacerEn {-100pt} {-90pt} \box110
\hfill
}
\vspace{-25pt}
\begin{fig}\label{the_join}
The pattern of a join-tile.
\end{fig}
}

\subsection{Synchronisation: the mechanisms}

The first mechanism which we introduce to force the synchronization of
different constriutions is that all bases on a given isocline merge. This
changes the tile for the coner, but this does not affect the algorithm
of section~5.

   The first principle forces the presence of various signals on the same
isocline. We have to look at the consequences of such simultaneous presence
in order to avoid contradictions which would ruin the construction.

   As a first point, note that the upper signals allow to differentiate the
various parts of a basis. If the upper horizontal signal is of the same colour
as the basis, we are outside any trilateral whose basis lies on the same 
isocline. If not, we are inside. We say that the basis is {\bf covered} if
it is accompanied by an upper horizontal signal of its colour. Otherwise, we
say that the basis is {\bf open}.

   This distinction is important. When a leg meets a basis: if it is the first  
half of a leg, {\it i.e.} between the vertex and the mid-point of the leg, it
meets the basis withouth changing it. When the second half of a leg meets a basis,
it crosses it if the colour is different. If it is the same colour, it crosses it 
only if the basis is covered. If it is open, the leg has met the basis with which
it forms a corner. Indeed, from lemma~\ref{unilateral}, an upper horizontal
signal cannot go from one leg of a trilateral to the other: there must be a
triangle of the same colour in between. And so, an open basis does not cross
the second half of a leg of a trilateral of the same colour: and so,
for the second half of a leg, when the leg meets such a signal, this means that 
the expected basis is found. 

   Now, we have to look at the consequences of the synchronization on other
signals: on the green signals and on the horizontal blue and red signals.

   The problem is the following. Consider two triangles~$A$ and~$B$ of the same
latitude. They belong to the same generation~$n$ and they have the same colour.
We may assume that~$A$ is on the left-hand side of~$B$. Let~$\iota$ be the
isocline of the mid-distance line of~$A$ and~$B$. From the study of the abstract
brackets, we know that there is a tower of phantoms inside~$A$ and inside~$B$, 
with also~$\iota$ as the mid-distance line of the phantoms of the tower. We also
know that the same tower is repeated along~$\iota$ between~$A$ and~$B$, possibly
several times. Accordingly, there is a green signal on~$\iota$ inside~$A$ and
inside~$B$, and also between~$A$ and~$B$.

   A similar problem occurs with the horizontal signals which are emitted by
the right-hand side leg of~$A$, the vertex being excepted. Symmetrically, the
left-hand side leg of~$B$ emits a horizontal signal of the same colour and on
the same isoclines of the latitude of~$A$. The join-tile of figure~\ref{the_join},
replicated in appropriate colours, allows to connect together horizontal signals
of opposite lateralities travelling between~$A$ and~$B$ in the appropriate
directions and on the same isocline. From lemma~\ref{unilateral}, if a horizontal 
signal enters a phantom of its colour, we get into trouble for the meeting with 
the other leg of the phantom.

   Both problems can be solved in a similar way.  

   The idea is to {\bf avoid} the phantoms as it is not possible to cross them.
The advantage is that the deviated signal does not distrub the construction
inside the phantom. Also, if during the detour, the signal keeps its laterality,
it behaves as if the phantom were not present. In particular, the join-tile can 
be used for the connection of the signal with the opposite one coming in front 
of it, from the other triangle. 

   Technically, the solution is the following. The legs of a triangle stop the
green signal which meet them on the mid-distance line, as required by the 
algorithm of section~5. Now, at the mid-point of the leg, on the corresponding 
isocline~$\iota$, the leg triggers an {\bf orange} signal~$\sigma$ of its 
laterality, outside the triangle, say~$A$. When $\sigma$ meets the first 
phantom~$P$ on its way, it does not cross it: on the ohter side of the leg
of~$P$, there is a green signal. Instead of this, $\sigma$ climbs up over the
left-hand side leg of~$P$ until it reaches the vertex. Then, it goes down along
the right-hand side leg of~$P$ until it reaches the isocline~$\sigma$. This is
easy to recognize: it is the single tile such that there is a green signal on 
the other side of the leg. Also, during this travel on the first half of 
the legs of~$P$, $\sigma$ does not change its laterality. From this, 
lemma~\ref{unilateral} also applies to~$\sigma$, which rules out the occurrence
of an orange signal on~$\iota$ inside~$P$. And so, $\sigma$ does not disturb the
construction inside~$P$. By lemma~\ref{unilateral}, we can see that 
$\sigma$~matches its opposite signal on~$\iota$ with the join-tile.

   It is not difficult to see that the solution applies to a horizontal blue
signal. Indeed, the structure of inner trilaterals inside a phantom is the same
as inside a triangle of the same generation. Accordingly, the notion of a free
row also applies to phantoms. Now, for a blue-0 or blue phantom, there is a
single free row: the mid-distane line. Accordingly, a horizontal blue signal 
meeting the leg of a blue-0 or blue phantom~$P$ on an isocline which
is not the mid-distance line of~$P$ also meets the opposite signal coming from
a blue-0 or blue triangle inside the phantom. The join-tile solves the problem 
if needed. For the blue signal which travels on the isocline~$\iota$ of the
mid-distance line of~$P$, its behaviour is exactly that of an orange signal, 
which solves the problem.

   We remain with the case of a red triangle. This time, $\iota$ still denotes
the mid-distance line of the triangles~$A$ and~$B$, and the right-hand side leg
of~$A$ emits right-hand side horizontal red signals on all the isoclines of the
latitude of~$A$, the line of the vertex of~$A$ being excepted. We have the 
same situation as with a blue signal if the signal arrives on an isocline
which does not correspond to a free row of the red phantom~$P$ which is 
first met on the way. If the signal arrives on an isocline of a free row 
of~$P$, the idea is to collect all the red signals travelling within the 
latitude of~$P$ on the isocline of a free row in a single signal, as there 
are a lot of free rows in a red trilateral. This single signal has the form 
of a red signal with a laterality: that of all the red signals arriving 
to~$P$. It climbs up over the legs of~$P$, this time from the vertex
to the basis and conversely. Also, the laterality of the signal is not 
changed and, when it goes down along the other leg of~$P$, the signal sends
a copy of itself, outside~$P$, on each isocline of a free row. In this way,
the red signals which arrive to~$P$ and which depart from it on an isocline 
of a free row of~$P$ constitue a comb: on one side, the comb gathers the signals,
and on the other side, it dispatches them. In this way, the avoiding is 
obtained without perturbing what happens inside~$P$, and also without 
disturbing what must be outside~$P$. The signal passes as if~$P$ were not 
present. We just remark that as the laterality is not changed, 
lemma~\ref{unilateral} also applies here. As a consequene, the same phenomenon
may happen inside~$P$, but only within the latitudes of the trianles which 
$P$~contains. As the mid-distance line of these triangles are different from
that of~$P$, the just described phenomenon occurs for the inner phantoms 
inside~$P$ whose mid-distance line is that of~$P$. This is conformal with 
the requirement that what happens inside~$P$ must not be disturbed.

   Now, we can conclude that the tiling forces the construction of trilaterals
generation after generation, as indicated by the algorithm os section~5.

\section{Completing the proof}

\subsection{The computing areas}

The {\bf active} seeds were defined in sub-section~6.1. They allow to define the
trilaterals in the hyperbolic plane.

   Now, we ignore the blue-0 and the blue triangles, the phantoms of any colour as 
well as the parts of the bases of red triangles which are covered. Accordingly,
we focus our attention on the red triangles only.

   We have already mentioned that lemma~\ref{unilateral} applied to horizontal
red signals allows us to detect the free rows inside the red triangles. We 
shall agree that ta special signal, the yellow one, without laterality, will
mark the free rows of the red triangles. In fact, the set of tiles which 
implement the interwoven triangles in the Euclidean plane, see
\cite{mmarXiv2,mmnewtechund}, also implement this detection and the installation 
of the yellow signal on the free rows.

   The free rows of the red triangles constitute the horizontals of the grid
which we construct in order to simulate the space-time diagram of a Turing
machine.

   Now, we have to define the verticals of the grid to complete the simulation.

   The verticals consist of rays which cross {\bf 8}-centres. 
Figure~\ref{vertical_1} illustrates how they are connected to the different
possible cases of contact of the isocline of free row with the border of the
tree. 

   The computing signal starts from a the seed. It travels on the free rows. 
Each time a vertical is met, which contains a symbol of the tape, the required 
instruction is performed. If the direction is not changed and the corresponding
border is not met, the signal goes on, on the same row. Otherwise, it goes down
along the vertical until it meets the next free row. There, it looks at the 
expected vertical, going in the appropriate direction. Further details ared 
dealt with in \cite{mmnewtechund}.

\vskip 15pt
\setbox110=\hbox{\epsfig{file=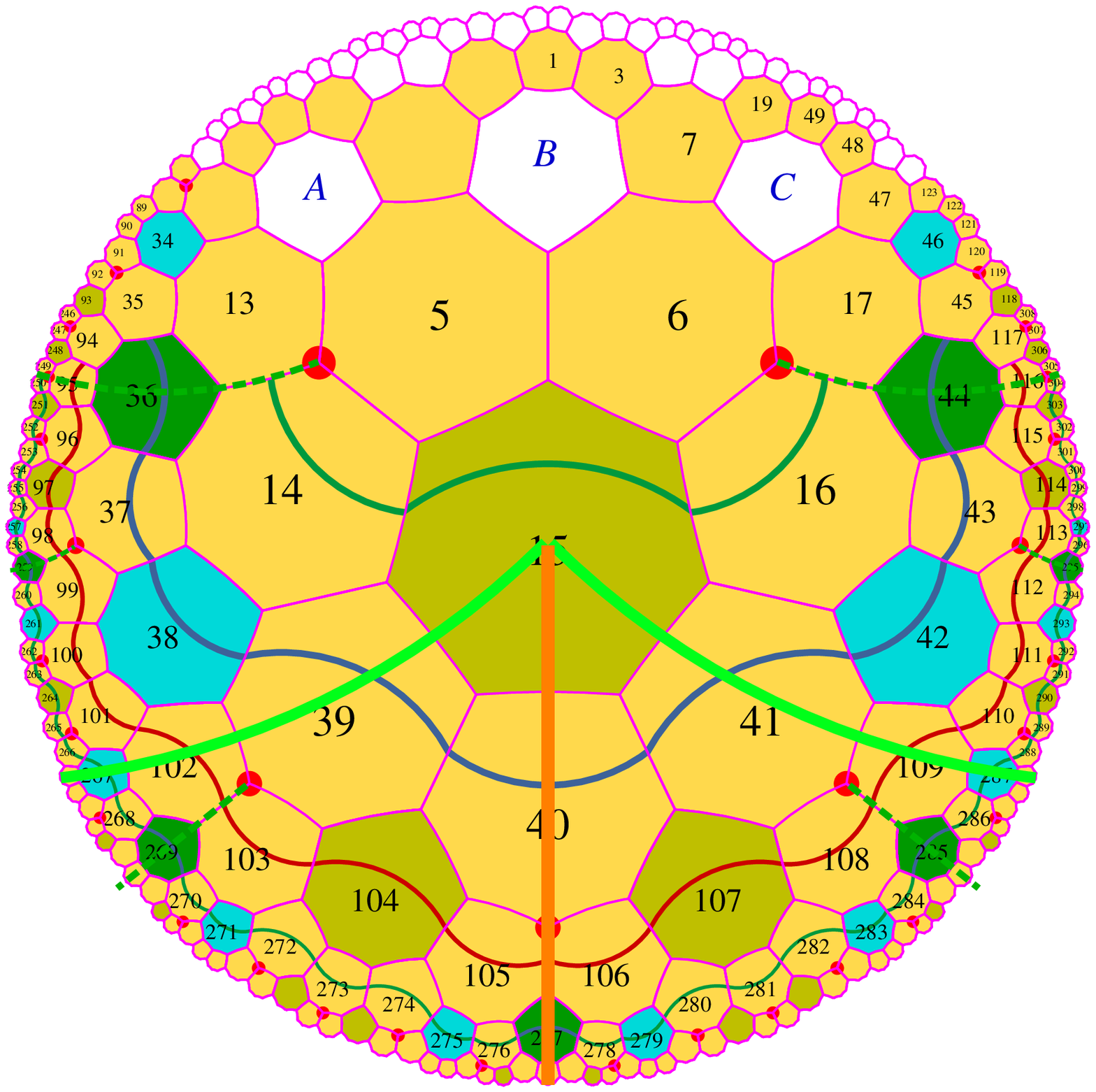,width=110pt}}
\setbox112=\hbox{\epsfig{file=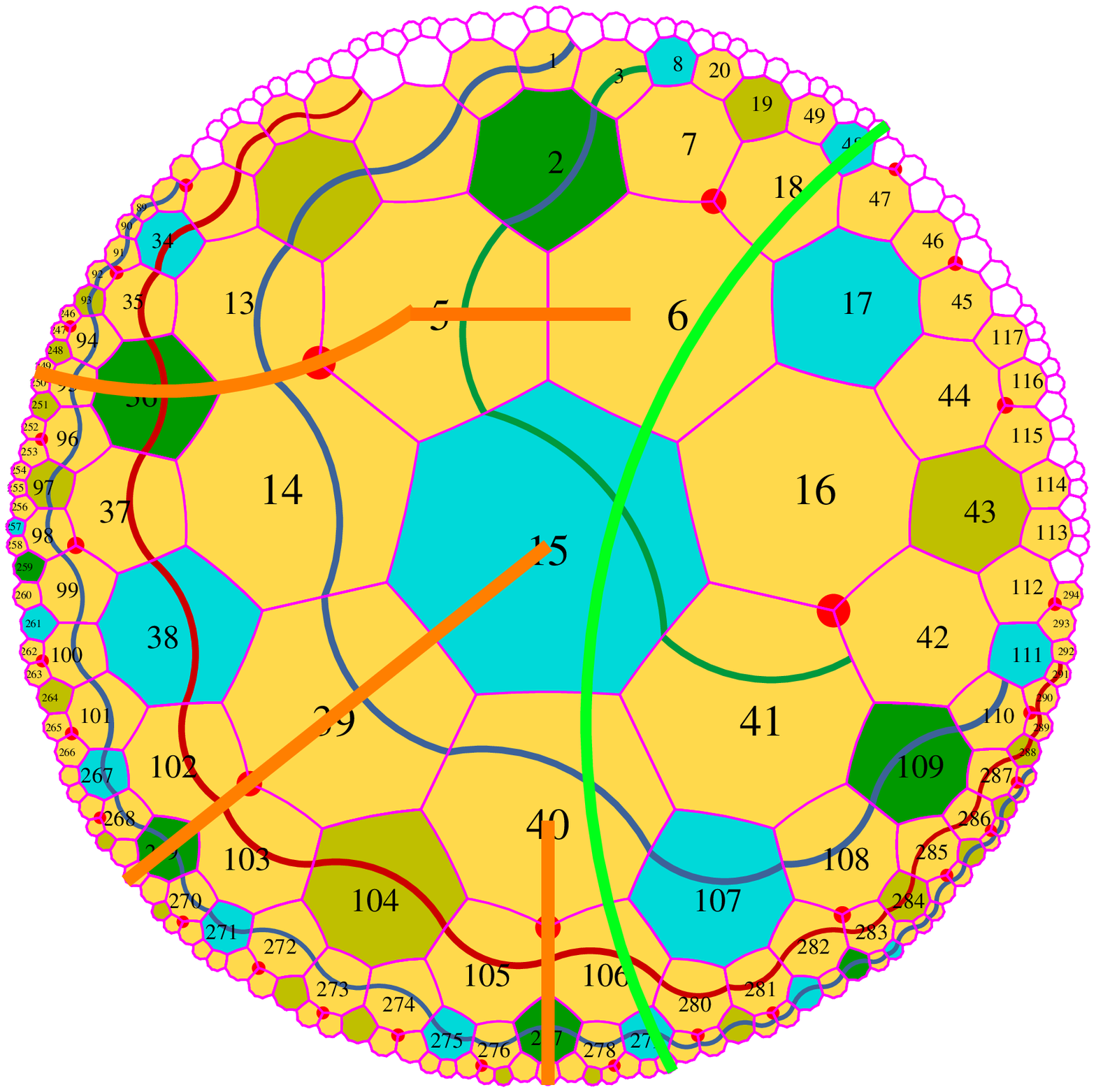,width=110pt}}
\ligne{\hfill
\PlacerEn {-130pt} {0pt} \box110
\PlacerEn {20pt} {0pt} \box112
\hfill}
\vskip-15pt
\begin{fig}\label{vertical_1}
\leurre
The perpendicular starting from a point of the border of a triangle which
represents a square of the Turing tape.
\vskip 0pt
On the left-hand side: the case of the vertex. On the right-hand side,
the three other cases for the right-hand side border are displayed on the
same figure.
\end{fig}

 Note that in the case of the butterfly model, see \cite{mmarXiv2,mmnewtechund},
the mechanism of the orange signal forces the green signal to run over the whole
isocline which is the mid-point of the latitude which contains no triangle. Indeed,
the laterality constraints of the tiles for the crossing legs of trilaterals
prevent an orange signal to run at infinity.

\subsection{The tiles}

   Within the frame of this paper, it is not possible to exhaustively describe
the tiles needed for the construction which we described. In the 
reports~\cite{mmnewtechund} and in~\cite{mmarXiv2}, we give a precise account
of the tiles.

   Here, we just indicate how to construct the tiles, at the same time giving a 
way to describe all the tiles and to count them.

   The finite set of tiles which we need to prove theorem~\ref{undec},
consists of two parts. First, we have the set of {\bf prototiles} which forces
the construction of the mantilla as well as the isoclines, the activation of seeds
through the isoclines~0 and the sreading of the scent, and then the construction 
of the interwoven triangles, including the detection of the free rows in the
red triangles.

   The second set consists of {\bf meta-tiles} which, in fact, are {\it variables}
for tiles, as the meta-tiles convey the signals directly connected with the
simulation of a given Turing machine. In the actual construction, the meta-tiles
replace a part of the prototiles: they replace all the prototiles which are
placed on an element of the computation: either the tiles which convey the 
computing signal, or those which convey the evolution of each square of the
Turing tape. But the replacement is not systematic: depeding on the simulated
machine, the same free row may hold a computing signal in a precise interval and 
no computing signal outside this interval.

   In each set, the tiles are constituted of a tile~$(\alpha)$ or~$(\beta)$
on which we superpose several signals. We distinguish between horizontal 
and vertical signals. The horizontal signals can be viewed as {\bf channels} 
which run along the path of the isocline inside the tile. We know that each tile
has a top part and a bottom one. By definition, the top is determined by the
side which is shared by the father of the tile. We define a {\bf local 
numbering} by numbering the sides of a tile from~1 up to~7, with the number~1
given to the side shared with the father, and the other numbers in increasing
order while counter-clockwise turing around the tile. In the local numbering,
the isocline goes from the side~2 to the side~7 in a white tile and it goes
from the side~3 to the side~7{} in a black one. For upper signals;the channel
is above the isoclines. A basis runs below the isocline. The green and orange 
signals run on the same channel which is above the isocline
and below the channel for a horizontal red or blue signal.

   For vertical signals, we have two main kinds of verticals: the legs of 
a trilateral and the verticals for the computing grid. The signals of the legs 
always cross black tiles and they are split into two categories: right-hand side
signals which go from the side~2 to the side~6 and left-hand side ones which go 
from the side~1 to the side~4. From this, we notice that a tile of a leg always 
knows its laterality and, also, which of its sides are inside the trilateral 
and which are outside it. The vertical signals for the grid go 
through~{\bf 8}-centres. In an {\bf 8}-centre, the signal goes from
the side~2 to the side~5 in terms of the local numbering. In terms of the
numbering of the mantilla, the signal goes from the side~7 to the
side~$\overline 4$. Next, in the petal $1\overline47\circ$, it goes from the
side~1 to the side~4 in terms of the local numbering. And then, in
the petal~$2\circ77$, it goes from the side~1 to the side~6, still with
respect to the local numbering.

   We just remark that we have three colours for the trilaterals: blue-0 for
the generation~0, blue for the even generations and red for the odd ones.
Legs of triangkes are represented by a {\bf thick} signal while legs of phantoms
are represented by a {\bf thin} one. First and second halves of legs are
also distinguished. In blue-0 and blue trilaterals, the first half is dark
and the second half is light. In red trilaterals, we use the opposite convention: 
the first half is light and the second one is dark. The tile of 
figure~\ref{the_join} defines the general patterns to indicate the laterality 
of a signal. These patterns are given the colour of the corresponding signal.

   Then, the most intricate task comes: the superposition of all these kinds
of signals. This can be precisely described and counted, as performed 
in \cite{mmnewtechund,mmarXiv2}.

  With this, we completed the proof of theorem~\ref{undec}.

\section{A few corollaries}

   The construction leading to the proof of theorem~\ref{undec} allows to get
a few results in the same line of problems.

   As indicated in \cite{goodmana,goodmanb}, there is a connection between~$GTP$
and the {\bf Heesch number} of a tiling. This number is defined as the maximum
number of {\bf coronas} of a disc which can be formed with the tiles of a given 
set of tiles, see \cite{mann} for more information. As indicated in \cite{goodmanb},
and as our construction fits in the case of domino tilings, we have the following 
corollary of theorem~\ref{undec}.

\begin{thm}\label{heesch}
There is no computable function which bounds the Heesch number for the tilings
of the hyperbolic plane.
\end{thm}

   The construction of~\cite{mmarXiv1,mmtechund} gives the following result,
see \cite{mmarXiv3,mmrp07}.

\begin{thm}\label{finite}
The finite tiling problem is undecidable for the hyperbolic plane.
\end{thm}

   Combined with the construction proving theorem~\ref{finite} and a 
result of~\cite{mmbook1}, the construction of the present paper allows us
to establish the following result, see~\cite{mmarXiv4}.

\begin{thm}\label{periodic}
The periodic tiling problem is undecidable for the hyperbolic plane, also
in its domino version.
\end{thm}

   In this statement, {\bf periodic} means that there is a shift which leaves
the tiling globally invariant.

   At last, in another direction, we may apply the argments of Hanf and Myers,
see~\cite{hanf,myers}, and prove the following result, see~\cite{mmnewtechund}.

\begin{thm}\label{nonrec}
There is a finite set of tiles such that it generates only non-recursive tilings
of the hyperbolic plane.
\end{thm}

\section{Conclusion}

   The first consequence is that, according to the estimations 
of~\cite{mmnewtechund,mmarXiv5}, we need a huge number of tiles. Taking into
account the changes introduced in~\cite{mmarXiv5}, a new counting indicates
that we need 23,323~tiles for the prototiles and 6,541~additional ones
for the meta-tiles, this will precisely be presented in a forthcoming paper.

   Of course, we may wonder whether the number of tiles can be reduced.
This might be possible by a small tuning of the present signals. As an example, 
we could forbid yellow rows on the mid-distance line of a red triangle. But
the advantage would not be important enough. Now, an attentive look at the
tables, suggested in sub-section~7.2, indicates that the reason of the big number
of tiles lies in lemma~\ref{iso5}. A consequence of the lemma is the important
number of passive tiles connected with the isoclines which do not bear the
construction signals. And so, to get a significant reduction of the number of
tiles means to find a new, simpler setting. Professor Goodman-Strauss advised me
to do so. In fact, recently, I realized that this is possible. I have not the 
room, here, to explain the idea. It is interesting to notice that this 
simplification allows to implement a similar implementation of the interwoven
triangles in both the ternary heptagrid and in the pentagrid.

   The second consequence which could be derived from the construcion lies
on a more abstract level. Let us look at the lifting of the abstract brackets
to the interwoven triangles. At first glace, this seems to be a Euclidean 
construction. In the very paper, a whole section is devoted to the Euclidean
implementation of the interwoven triangles. And in the next section, we still
transfer this construction to the hyperbolic plane. It seems to me that the fact
that this transfer is possible has an important meaning. From my humble point
of view, it means that a construction which looks like purely Euclidean has
indeed a purely combinatoric character. It belongs to absolute geometry and it
mainly requires Archimedes' axiom. Note that absolute geometry itself has no
pure model. A realization is necessarily either Euclidean or hyperbolic. We
suggest to conclude that, probably, the extent of aboslute geometry is
somehow under-estimated. 

\section*{Acknowledgement}

I am very pleased to acknowledge the interest of several colleagues and
friends to the main result of this paper. Let me especially thank Andr\'e
Barb\'e, Jean-Paul Delahaye, Chaim Goodman-Strauss, Serge Grigorieff, Tero 
Harju, Oscar Ibarra, Hermann Maurer, Gheorghe P\u aun, Grzegorz Rozenberg
and Klaus Sutner.

\end{document}